\documentclass[prl,preprint]{revtex4}
\usepackage[utf8]{inputenc}

\usepackage{amsmath}
\usepackage{amssymb}
\usepackage{natbib}
\usepackage{graphicx}
\usepackage[usenames,dvipsnames]{xcolor}
\usepackage{siunitx}
\usepackage{xfrac}
\usepackage[utf8]{inputenc}
\usepackage{braket}
\usepackage{soul}

\colorlet{faintblue}{blue!50!white}
\colorlet{darkgreen}{green!30!black}
\colorlet{purple}{magenta!50!black}

\newcommand{\wavenumber}{\per\cm}
\newcommand{\mat}[1]{\mathbf{#1}}
\renewcommand{\vec}[1]{\mathbf{#1}}

\newcommand{\paren}[1]{\left(#1\right)}
\newcommand{\KG}{\Gamma}
\newcommand{\KK}{\mathrm{K}}
\newcommand{\KM}{\mathrm{M}}

\usepackage[nopostdot,nogroupskip,nonumberlist]{glossaries}

\newacronym{FBZ}{FBZ}{first Brillouin zone}
\newacronym{LO}{LO}{longitudinal optical}
\newacronym{CG}{CG}{conjugate gradient}
\newacronym{LA}{LA}{longitudinal acoustic}
\newacronym{ZO}{ZO}{out-of-plane optical}
\newacronym{ZA}{ZA}{out-of-plane acoustic}
\newacronym{TO}{TO}{in-plane transverse optical}
\newacronym{TA}{TA}{in-plane transverse acoustic}
\newacronym{FK}{FK}{Frenkel-Kontorova}
\newacronym{NFE}{NFE}{nearly-free electron}
\newacronym{NFP}{NFP}{nearly-free phonon}
\newacronym{KC}{KC}{Kolmogorov--Crespi}

\newcommand{\translationOper}[1]{\hat{T}\!\paren{#1}}

\begin{document}
\title{Soliton signature in the phonon spectrum of twisted bilayer graphene}
\author{Michael Lamparski}
\affiliation{Department of Physics, Applied Physics, and Astronomy, Rensselaer Polytechnic Institute, Troy, New York 12180, United States}
\author{Benoit Van Troeye}
\affiliation{Department of Physics, Applied Physics, and Astronomy, Rensselaer Polytechnic Institute, Troy, New York 12180, United States}
\author{Vincent Meunier}
\affiliation{Department of Physics, Applied Physics, and Astronomy, Rensselaer Polytechnic Institute, Troy, New York 12180, United States}

\begin{abstract}
The phonon spectra of twisted bilayer graphene (tBLG) are analyzed for a series of 692 twisting angle values in the $[0,\SI{30}{\degree}]$ range. The evolution of the phonon bandstructure as a function of twist angle is examined using a band unfolding scheme where the large number of phonon modes computed at the $\Gamma$ point for the large moir\'e tBLG supercells are unfolded onto the Brillouin Zone (BZ) of one of the two constituent layers. In addition to changes to the low-frequency breathing and shear modes, a series of well-defined side-bands around high-symmetry points of the extended BZ emerge due to the twist angle-dependent structural relaxation. The results are rationalized by introducing a nearly-free-phonon model that highlights the central role played by solitons in the description of the new phonon branches, which are particularly pronounced for structures with small twist angles, below a buckling angle $\theta_{\rm B}\sim\SI{3.75}{\degree}$.
\end{abstract}

\maketitle
Unconventional superconductivity and magnetism were not typically associated with graphene until very recently with the observation of strong electronic correlation in bilayer graphene twisted roughly at the $\theta=\SI{1.1}{\degree}$ \textit{magic} angle~\cite{Cao2018,Cao2018b,Yankowitz2019,Sharpe2019,Codecidoeaaw2019}.  This effect is related to the emergence of flat electronic bands with narrow band width~\cite{Cao2018,Codecidoeaaw2019,Nam2017,Uchida2014,Guinea2019}, and theoretical studies have established the importance of structural relaxation in the description of these phenomena~\cite{Alden2013,Nam2017,Wijk2015,Dai2016,Jain2016,Choi2018,Koshino2019b,Yong2019}. The atomic relaxation in twisted bilayer graphene (tBLG) is linked to the development of strain fields that tend to minimize the unfavorable AA-stacking regions---with local buckling---and lead to the formation of triangular domain patterns in the moir\'e pattern. This structural re-arrangement can be interpreted as the formation of solitons in the system~\cite{Frank1949,Popov2011, Alden2013}.

Most of the existing theoretical studies devoted to tBLGs have focused on electronic properties. However, tBLG phonons are also expected to be affected by the formation of the moir\'e superlattice and its underlying strain fields, since they fundamentally correspond to excitations of the lattice~\cite{Cocemasov2013,Choi2018,Angeli2019}. Studying the large unit-cells corresponding to small values of twisting angles is computationally challenging and a fine analysis of the tBLG phonons has proven difficult, since a full treatment requires the calculations of dynamical matrices with hundreds of thousand elements for structures prepared at a minimum energy configuration. Angeli and co-workers~\cite{Angeli2019} were able to identify 10 nearly-flat energy phonon bands---not present in single-  or bi-layer (Bernal) graphene---among the $\sim 30,000$ phonon modes of one specific tBLG. At the same time, continuum models have predicted band gap openings at the zone border of the tBLG reciprocal lattices~\cite{Koshino2019a,Koshino2019b}. These studies are promising but need additional refinements to include buckling.

In this Letter, we overcome the technical difficulty of  analyzing the large number of phonon modes of tBLG by adopting an unfolding scheme~\cite{allen2013RecoveringHiddenBloch}. The many phonon modes of the superlattice are unfolded onto the reciprocal lattice of the bottom layer of graphene. This allows us to demonstrate the existence of a rich phononic structure, especially for small twist angles. The present study not only confirms the emergence of new phonon side-bands at the high-symmetry points of the graphene reciprocal lattice, but it also predicts a number of features that are specific to tBLGs. In particular, our study predicts very significant band splitting as large as $\SI{50}{\wavenumber}$ for the high-frequency $G$ modes at the zone center of structures with small twist angles. A careful monitoring of the high-symmetry phonon modes as a function of the twist angle reveals the elaborate, but continuous, evolution of these phonon side-bands for $\theta \rightarrow 0$. Notably, the layer breathing mode at the zone center splits into several bands for decreasing angle---some showing evanescent character, others featuring mixed in-plane and out-of-plane characters.

A large number of tBLG superlattices are constructed following a procedure similar to the one presented in Ref.~\onlinecite{shallcross2010}.  Certain discrete twist angles produce a commensuration between the layers, described by the equation $\mat C_B \mat B = \mat L = \mat C_T \mat T$, where $\mat B$ and $\mat T$ are matrices whose rows are basis vectors for each layer's lattice (bottom and top), $\mat L$ is the same for the commensuration cell, and $\mat C_B$ and $\mat C_T$ are integer matrices. Starting with this equation and applying a number of basic number theoretical arguments, a one-to-one correspondence can be found between twist angles $\theta \in \left[\SI{0}{\degree}, \SI{90}{\degree}\right]$ and solution triplets $(a, b, c)$ to the Diophantine equation $a^2 + 3b^2 = c^2$, where $a$, $b$, and $c$ are mutually coprime, non-negative integers. These solutions have twist angles of $\cos \theta=a/c$ and contain $N=4c/\gcd(2, c)$ atoms per unit cell, where gcd is the greatest common divisor function. Of these solutions, only those with twist angles in the range $\SI{0}{\degree} \leq \theta < \SI{30}{\degree}$ are worth considering due to symmetry. There are in principle an infinitude of angles that lead to a commensurate structure, but only 692 of these angles produce a structure containing fewer than 20,000 atoms. Those are the systems we investigate in this work.

We use classical force-fields for structural relaxation and to compute the corresponding phonons in the harmonic approximation. Intralayer forces are computed using the second-generation REBO potential~\cite{brenner2002SecondgenerationReactiveEmpirical}, while interlayer forces are modeled using the registry-dependent \gls{KC} potential~\cite{kolmogorov2005RegistrydependentInterlayerPotential}, in its local normal formulation. All atomic positions and lattice parameters of the superlattice are optimized with a \gls{CG} algorithm. The latter are found to vary in a negligible manner compared to the rigidly-stacked case.  After CG convergence of all force components within $\SI{e-3}{\eV\per atom}$, the dynamical matrix at $\Gamma$ is computed using finite differences and then diagonalized. The presence of many local extrema makes the accurate relaxation of tBLG tedious. However, we take advantage of the phonon calculation to operate an iterative assessment of dynamical stability. In practice, when modes are found with negative eigenvalues (complex frequencies) and if they are not translational acoustic modes, CG is performed again in a reduced parameter space where each parameter corresponds to one of the negative modes. This accelerated procedure can in theory be repeated until all negative modes are removed from the system. Unfortunately, suppressing the entirety of the negative modes for the small twist angle structures has been found to be particularly challenging ($>$11 iterations required); however, the imaginary frequencies of these modes have a magnitude $<\SI{4}{\wavenumber}$ and are thus assumed to have negligible influence on the oncoming discussion.

Similar to previous studies~\cite{Alden2013,Nam2017,Wijk2015,Dai2016,Jain2016,Choi2018,Koshino2019b,Yong2019}, accommodation between the layers is observed in order to minimize the AA-stacking region, which proves especially significant for small twist angles. Buckling is found to occur for certain structures with twist angles that are below some critical \textit{buckling angle} $\theta_{\rm B}$, defined here as the angle below which negative modes are observed for the corresponding structure after one full CG relaxation. This definition accounts for the fact these non-zero modes are responsible for the buckling of the system.  Here, we find $\theta_{\rm B}\sim\SI{3.75}{\degree}$, meaning that all structures below this angle buckle, with the notable exception of a few structures (e.g. $\SI{3.482}{\degree}$ and $\SI{3.150}{\degree}$, with $(a,b,c)=(541, 19, 542)$ and $(661, 21, 662)$, respectively) that have particularly small unit cells.

\begin{figure*}[htp]
 \includegraphics[width=0.9\textwidth]{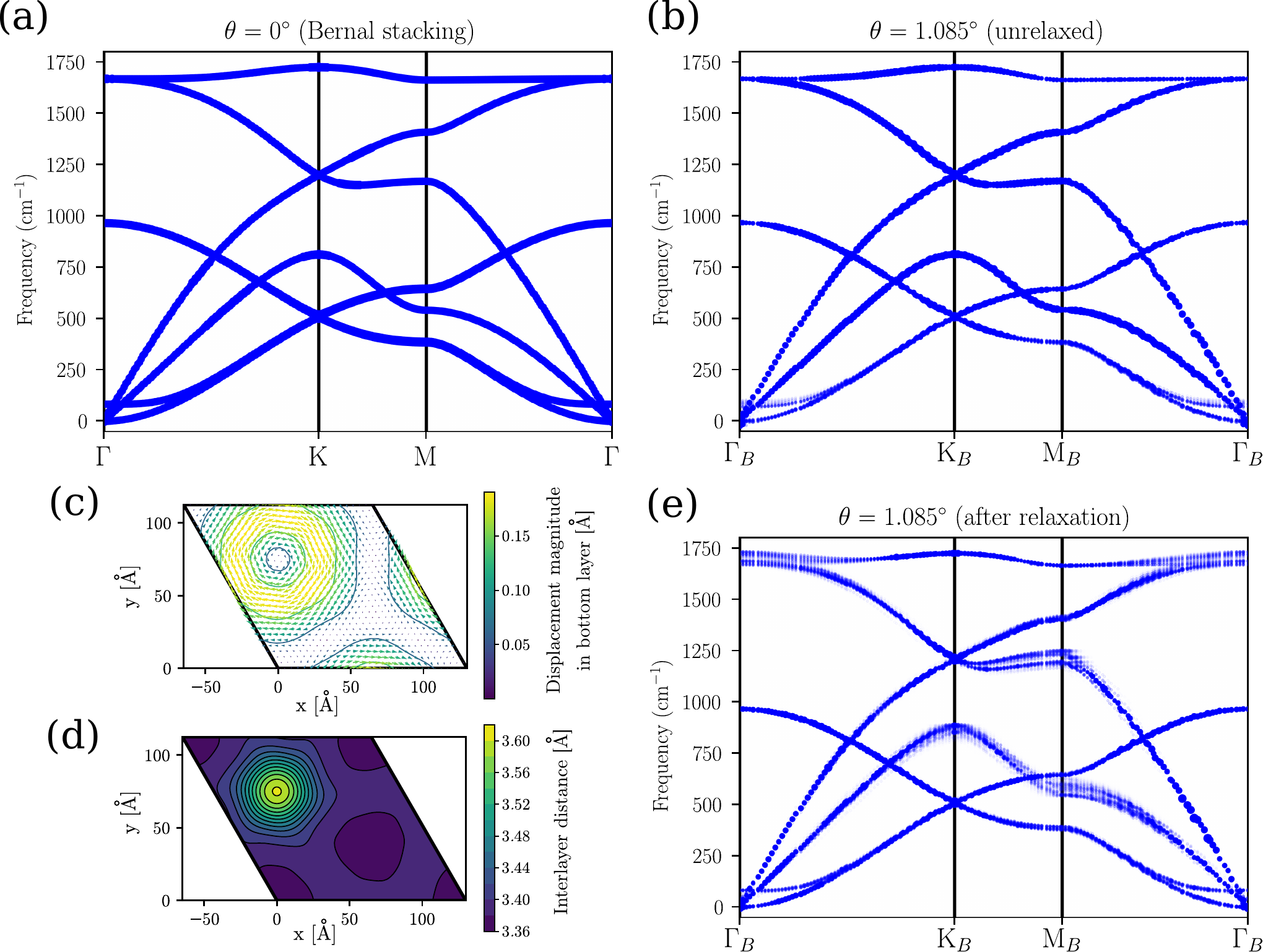}
 \caption{\label{fig:unfold-bands}  (a) Phonon band structure of graphene bilayer in the Bernal stacking configuration. (b) Unfolded phonon band structure of twisted bilayer graphene structure with rotation angle $\SI{1.085}{\degree}$ prior to relaxation.  The $B$ subscripts denote that the high symmetry path is taken along the Brillouin zone of the bottom layer, and each position on the x-axis displays data unfolded onto the nearest point that is an image of $\Gamma_L$, $\KM_L$, or $\KK_L$. (c) In-plane displacement field in the bottom layer of the $\SI{1.085}{\degree}$ twist-angle structure. Close to the AA-stacking region, this layer tends to rotate clockwise in order to minimize the region of AA stacking. (d) Variation of the interlayer distance for the $\SI{1.085}{\degree}$ twist-angle structure. The most pronounced variations are observed close the AA-stacking region, indicating local buckling of the layers. (e) Unfolded phonon band structure of the  $\SI{1.085}{\degree}$ twist-angle structure after relaxation. Compared to the unrelaxed case (panel (c)), significant splitting can be seen at $\KM$, $\KK$, and in the G band.}
\end{figure*}

As a reference, Fig.~\ref{fig:unfold-bands}(a) represents the phonon band structure of Bernal-stacked bilayer graphene as computed with the REBO+\gls{KC} potential.  This potential has difficulty describing the dispersion of the high-frequency branch~\cite{Maultzsch2004}, but all of the other branches of graphene are qualitatively reproduced. Note that it also appears to somewhat underestimate the strength of the interlayer interactions: the shear mode for AB-stacked graphene is found to be $\SI{19.7}{\wavenumber}$ (\textit{versus} $\SI{32}{\wavenumber}$ from experiment~\cite{tan2012ShearModeMultilayer}), and the layer-breathing mode in structures around $\SI{12}{\degree}$ is found to be $\sim\SI{79}{\wavenumber}$ (\textit{versus} $\sim\SI{95}{\wavenumber}$ from experiment~\cite{he2013ObservationLowEnergy}).

Analyzing the effect of twist angle on the phonon modes in tBLGs is particularly difficult since each structure has a very large number ($3N$) of normal modes, and this number varies among all structures considered. To address this issue, the phonon data calculated at the supercell zone center $\KG_L$ (i.e., $\Gamma$ in the small BZ of the tBLG) are unfolded onto the BZ of reciprocal cell of the primitive cell of one of the layers, producing a band structure in the single-layer \gls{FBZ}~\cite{allen2013RecoveringHiddenBloch}. This process is \textit{fuzzy} in the sense that a mode in the supercell is usually expressed a linear combination of single-layer modes with different wavevectors. Note also that the images of $\Gamma_L$ in the reciprocal cell of one layer are not compatible with the translational symmetry of the other layer; thus, the unfolding process exclusively considers the projections of the normal modes onto a single layer.  Applying the unfolding method requires the construction of translation operators $\translationOper{R}$ for $R$ in the quotient group $\mathbf{L}_B/\mathbf{L}_L$, where $\mathbf{L}_B$ is the real-space lattice of the bottom layer, and $\mathbf{L}_L$ is the moir\'e lattice. Ideally, we apply these operators to a Bloch function represented in the form of a $3N$-dimensional vector $\vec u_{\vec K}$ (for some $\vec K$ in the moir\'e \gls{FBZ}), containing Cartesian components at each site in the moir\'e cell.

Prior to relaxation, it can be seen that each translation operator may take the form $\translationOper{R} \vec u_\vec K = \mat Q_{\vec K, \vec R}\mat P_{\vec R} \vec u_\vec K $, where $\mat P_{\vec R}$ is a permutation matrix (obtained by recognizing that the translation is a symmetry of the atomic coordinates in this layer), and $\mat Q_{\vec K, \vec R}$ is a diagonal matrix of corrective phase factors (fixing the Bloch phases for atoms that are translated between different images of the moiré cell). Strictly speaking, after relaxation, these quotient group translations are no longer symmetries of the atomic coordinates within the layer.  However, we may continue to use the above result by taking as an approximation that the atoms vibrate around their unrelaxed positions for the purpose of unfolding. This assumption is violated somewhat in small angle structures, especially due to out-of-plane buckling. However, this out-of-plane motion is not expected to meaningfully affect the unfolding probabilities onto points that lie in the plane.

\begin{figure*}[htp]
 \includegraphics[width=0.9\textwidth]{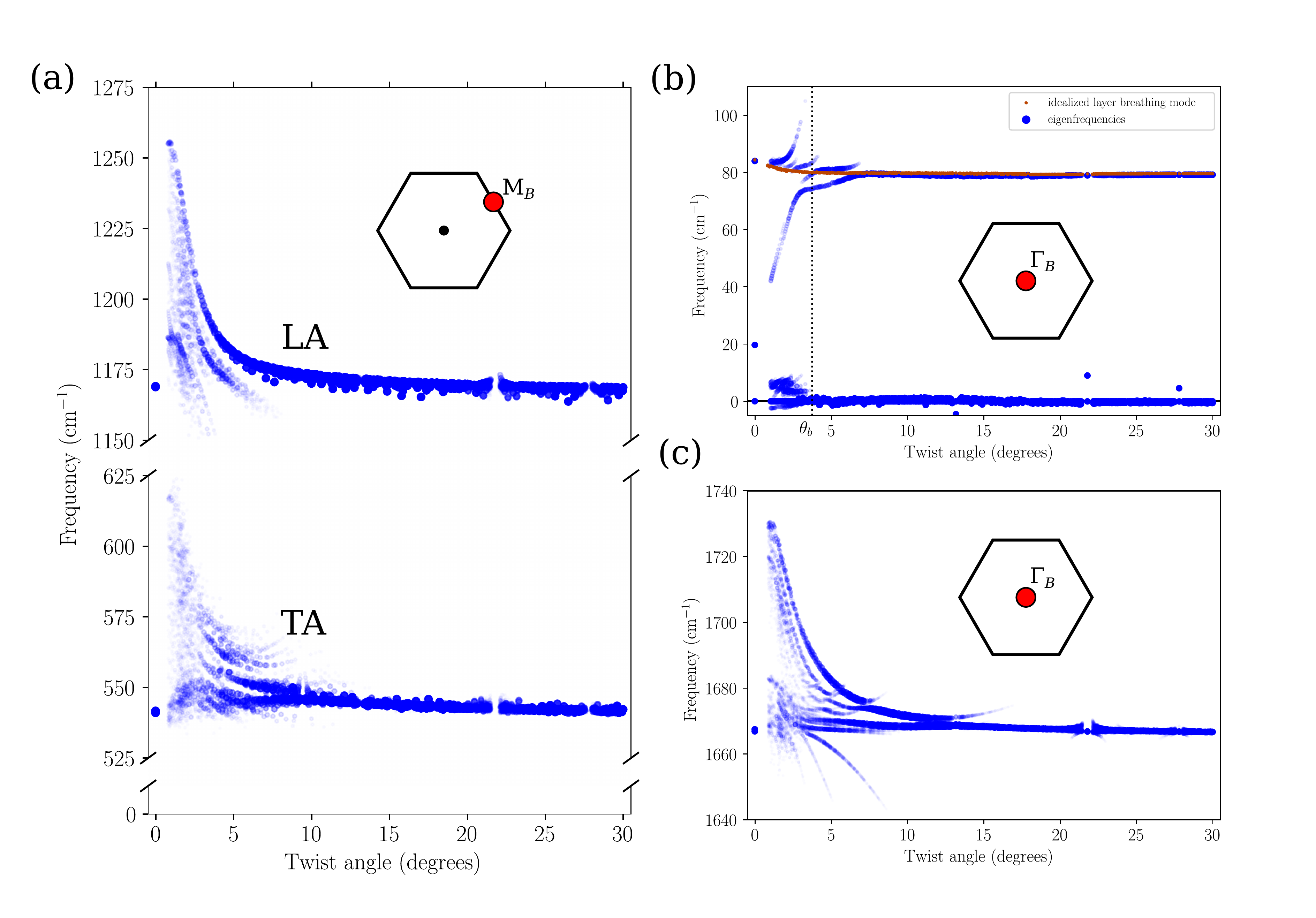}
 \caption{\label{fig:unfold-points}   (a) Phonon bands unfolded from the moir\'e $\Gamma$ point onto the single-layer M point at a variety of twist angles.  Of the 692 tBLG structures with under 20000 atoms per cell, only 617 of them have an image of $\Gamma_L$ that lies within $0.01\times \smash{2\pi\si{\per\angstrom}}$ from $\KM_B$; the rest are not shown (except for the data at $\SI{0}{\degree}$, which is computed without unfolding).  (b-c) Phonon bands unfolded onto the single-layer $\Gamma$ point for all 692 structures at (b) low and (c) high frequencies (i.e., including graphene's G band).  The low frequency plot includes the frequencies derived from the expectation value of a pure layer-breathing mode (red). The lower frequency values correspond to shear-like modes.   Structures below $\theta_B=\SI{3.75}{\degree}$ are found after CG to have buckling modes that produce more energetically favorable structures. Relaxation is performed along these modes up to three times.}
\end{figure*}

We find that even at a moderately-sized angle of $\SI{9.737}{\degree}$  (not shown here), the unfolded band structure exhibits no features significantly different from those of Bernal stacking, except for the shear mode frequency (which is much lower than that of Bernal bilayer graphene due to a reduced effective interlayer force constant).  In stark contrast, at small nonzero twist angles, several additional prominent features appear in the unfolded bands. For instance, Fig.~\ref{fig:unfold-bands}(e) depicts the unfolded representation for the phonons of the $\SI{1.085}{\degree}$ ($(a,b,c)=(5581, 61, 5582)$) twist-angle structure. This phonon band structure exhibits the basic features of the AB-stacked bilayer graphene, including the prominent splitting of the \gls{ZA} band to produce a layer-breathing mode. Further, there is splitting of the G band at the single-layer $\Gamma$, by about $\SI{60}{\wavenumber}$ at $\theta=\SI{1.085}{\degree}$.  In addition to this splitting, branches form in a number of side-bands around the single-layer M and K points.   Structural relaxation plays a critical role in all of these effects; for comparison, band plots are included for structures with and without relaxation in Fig.~\ref{fig:unfold-bands}(e) and Fig.~\ref{fig:unfold-bands}(b), respectively. A plot of the in-plane displacement fields minimizing the AA-stacking region is shown in Fig.~\ref{fig:unfold-bands}(c), while the variation of the interlayer distance is provided in Fig.~\ref{fig:unfold-bands}(d) to highlight the buckling in the vicinity of the AA-stacking region. Clearly, without relaxation, none of the side band effects are apparent, and the phonon band structure is extremely similar to the case of Bernal graphene bilayer, except for a reduced shear mode frequency.

The plots in Fig.~\ref{fig:unfold-points} focus on specific high-symmetry points in the single-layer Brillouin Zone and compare the unfolded bands between different structures.  Fig.~\ref{fig:unfold-points} (a) shows the evolution of the bands at $\KM_B$ with respect to changing angle.  In this plot, only data from $\Gamma_L$ point is used.  $\KM_B$ is not an image of $\Gamma_L$ under the moir\'e reciprocal lattice; thus, for each structure, the data at the nearest image of $\Gamma_L$ is used.  For many of these structures (particularly those with smaller commensurate cells), this image may be far away from $\KM_B$, resulting in poor-quality data; therefore, we omit structures where the nearest image of $\Gamma_L$ is further than $0.01\times \smash{2\pi\si{\per\angstrom}}$ from $\KM_B$.  The \gls{TA} and \gls{LA} bands can be observed to split into increasingly many side bands as angle decreases (though there is no splitting at $\SI{0}{\degree}$).

Unfolded probability data at $\Gamma_B$ are available for all 692 structures, and are shown in Fig.~\ref{fig:unfold-points}(b-c).  Astonishingly, all the modes do not undergo a continuous transformation from $\SI{30}{\degree}$ to $\SI{0}{\degree}$.  Perhaps most strikingly is what happens to the layer breathing mode (Fig.~\ref{fig:unfold-points}(b) around $\SI{80}{\wavenumber}$), which is well defined at large angles but splits into several branches for decreasing angles. One of them shows a linear dependency that can be extrapolated onto the shear mode at $\SI{0}{\degree}$, implying that this mode shows both in-plane and out-of-plane characters for small twist angles. The others are evanescent and show complex behavior as a function of twist angle. Nevertheless, their general trend seems to follow the expectation value of the dynamical matrix for the layer breathing mode in the system (i.e., matrix element of the dynamical matrix for a pure breathing mode; red curve in Fig.~\ref{fig:unfold-points}(b)), and appears to eventually converge to the value of the layer breathing mode in Bernal graphene bilayer for $\theta\to 0^\circ$.  

Finally, the high-frequency modes at the tBLG zone center are represented in Fig.~\ref{fig:unfold-points}(c). Similar to the modes shown in Fig.~\ref{fig:unfold-points}(a), more and more side bands emerge for decreasing angles. Notably, there are several branches that take values higher than that of the $G$ mode computed in Bernal graphene bilayer. It should be emphasized that this is the most striking signature of twisting angle and it could, in principle, be used as a fingerprint of a specific twisting angle. 

 \begin{figure*}[htp]
 \includegraphics[width=0.9\textwidth]{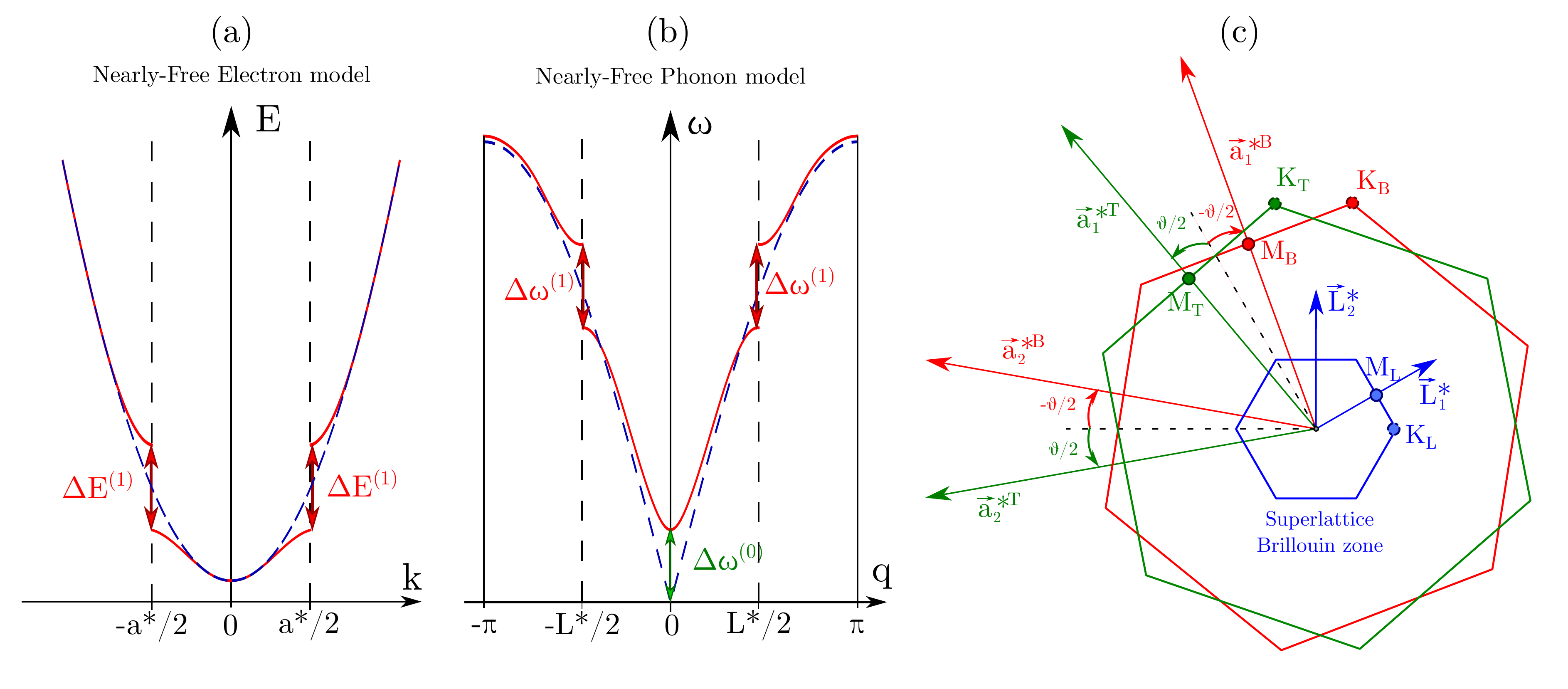}
 \caption{\label{fig:model}  (a-b) Schematic representation of the predictions of the NFE and NFP models, truncated to the first potential harmonic for the sake of legibility. The dashed blue lines represent the unperturbed spectra, \textit{i.e.} the free electrons and free phonons, while the red lines correspond to the perturbed solutions. Both models predict a band gap opening either at the lattice Brillouin zone $a^*/2$ (NFE model) or at the superlattice Brillouin zone $L^*/2$ (NFP model). In addition, the NFP model leads to the transformation of the acoustic phonon into an optical one. (Right) Schematic representation of the Brillouin zones of the rotated graphene lattice and of the superlattice. Due to the geometry of the problem, a high-symmetry point lying on the superlattice Brillouin zone corresponds exactly to the difference in terms of position between the corresponding high-symmetry points in the bottom and top layers (e.g. $\KM_L$, $\KM_B$ and $\KM_T$).}
 \end{figure*}
 
We will now discuss the fundamental origin of the tBLG-specific phonon side bands, starting from the \gls{FK} model~\cite{Frank1949, Braun1998}. This classical model considers a 1D chain of atoms placed into a periodic external potential whose period differs from the one of the chain. In general, its ground state consists of the periodic repetition of solitons that separate regions where the atoms sit in the close neighborhood of potential minima~\cite{Frank1949}. Those solitons can be interpreted as new quasi-particles of the system, and they behave relativistically since their dynamics are dictated by the Sine-Gordon equation~\cite{Braun1998}. Note that in addition to a soliton solution, the formal problem admits another type of quasi-particle solution called \textit{breathers}. These consist of non-linear waves oscillating in time but localized in space~\cite{Braun1998}.

Investigating the dynamics of breathers and solitons is a problem of interest on its own. However, we shall here focus on examining how their static properties affect the phonons. The soliton network deforms the lattice periodically, generating a potential that is felt by the otherwise-free (i.e., independent) phonons. Let us assume that this generated potential is weak and thus can be treated perturbatively.  This problem strongly resembles the one described by the \gls{NFE} model~\cite{Rabe2002}, except that the electrons and ionic potential are replaced by the phonons and the soliton potential, respectively. Insights into the phonons under the soliton potential, hereby referred to as the \gls{NFP} model, can therefore be extracted from the \gls{NFE} model.

The \gls{NFE} model accounts for the opening of band gaps in the electronic band structure, located at the zone borders for the odd harmonics of the ionic potential and at the zone center for the even harmonics (see Fig.~\ref{fig:model}(a)). The mean value of the potential is generally discarded since it simply leads to a rigid energy shift. Restricting the analysis to the first harmonic of the potential, the band gap opening at zone borders is associated with the formation of stationary waves: the electrons are either localized around the nuclei (lower energy state) or away from them (upper energy state).

The results for phonons in the \gls{NFP} model are unsurprisingly similar to the ones derived in the \gls{NFE} model for electrons with the emergence of band gaps associated with stationary waves involving the displacement (or pinning) of the soliton cores. The full resolution will be provided elsewhere but the main results are shown in Fig.~\ref{fig:model}(b). A number of differences with respect to the \gls{NFE} model should be noted. First, the spectra of free electrons and free phonons differ: the former is unbound ($E\propto k^2$, where $E$ is the energy and $k$ the electron wavenumber) while the latter is bound ($\omega^2 \propto \sin^2{(q/2)}$, where $\omega$ is the frequency and $q$ the phonon wavenumber). This translates into noticeable differences in the dispersion relations of the two nearly-free models. Second and more fundamentally, the acoustic mode present in the isolated 1D chain is transformed into an optical mode in the NFP model ($\Delta \omega^{(0)}$ opening in Fig.~\ref{fig:model}). This comes from the fact that the mean value of the potential cannot be discarded in the case of phonons since the soliton perturbation is acting over the square of the frequency. This contrasts with the NFE model, where the perturbation acts directly on the energy and could therefore be disposed of. 

The emergence of phonon side bands instead of band gaps as predicted by the \gls{NFP} model is a particularity of the twisted bilayer system. To understand this, it is useful to inspect the BZ of the constituent layers with respect to that of the superlattice (see Fig.~\ref{fig:model}(c)). By geometrical construction, the superlattice reciprocal vectors correspond exactly to the difference between the reciprocal vectors of the bottom and top layers~\cite{Nam2017,Koshino2019a}. As a consequence, the phonons at the $\KM_L$ superlattice high-symmetry points correspond to phonon states at the $\KM_B$ and $\KM_T$ high-symmetry points in the bottom and top layers, respectively. The same reasoning holds for the $\KK_L$ high-symmetry point. As discussed before, the \gls{NFP} model indicates that a band gap is expected to open at the high-symmetry points of the superlattice Brillouin zone. It follows that the first soliton potential harmonic opens a band gap at the $\KM_L$ point with amplitude $\Delta\omega^{(1)}$ and the third harmonic opens a band gap of amplitude $\Delta \omega^{(3)}$ at the same point, etc. The mechanism hinges on the fact that, in contrast to electronic potential, the dispersion relation of phonon is bound: the band gap openings always affect the same initial phonon state, leading to the creation of different phonon side bands around the same frequency window. This explains the origin of the numerous phonon side bands presented in Fig.~\ref{fig:unfold-bands}, which can thus be understood as the intrinsic signature of solitons in the system. 

To summarize, we find the rich phononic structure for small twist angles in tBLGs is closely associated with the formation of a soliton network and can be explained in terms of the NFP model introduced here. Fundamentally, compared to the NFE model, the role of electrons and nuclear potential are fulfilled by the phonons and soliton potentials in the NFP model, respectively. The general framework allows us to contrast the results of the NFP and NFE models, in term of the different dispersion relations adopted by the phonons and electrons. 

One should mention that, while we chose the single-layer \gls{FBZ} as a support for unfolding, the mechanism of band unfolding is not unique and can be applied to other choices of BZ. For this reason, a number of bands cannot be directly seen in plots such as those presented in Fig. 1. For example, the R bands described by Jorio et al. in Ref.~\cite{jorio2013RamanSpectroscopyTwisted} do not lie along the high symmetry path chosen here but can be obtained in a way similar to the procedure described in this Letter. This effect is outside of the scope of the present study and will be discussed in details elsewhere.

\end{document}